# Phonon spectra in CaFe$_2$As$_2$ and Ca$_{0.6}$Na$_{0.4}$Fe$_2$As$_2$: Measurement of the pressure and temperature dependence and comparison with ab-initio and shell model calculations

R. Mittal[1,2], S. Rols[3], M. Zbiri[3], Y. Su[1], H. Schober[3], S. L. Chaplot[2], M. Johnson[3], M. Tegel[4], T. Chatterji[5], S. Matsuishi[6], H. Hosono[6], D. Johrendt[4] and Th. Brueckel[1,7]

[1]*Juelich Centre for Neutron Science, IFF, Forschungszentrum Juelich, Outstation at FRM II, Lichtenbergstr. 1, D-85747 Garching, Germany*

[2]*Solid State Physics Division, Bhabha Atomic Research Centre,Trombay, Mumbai 400 085, India*

[3]*Institut Laue-Langevin, BP 156, 38042 Grenoble Cedex 9, France*

[4]*Department Chemie und Biochemie, Ludwig-Maximilians-Universitaet Muenchen, Butenandtstrasse 5-13 (Haus D), D-81377 Muenchen, Germany*

[5]*Juelich Centre for Neutron Science, Forschungszentrum Juelich, Outstation at Institut Laue-Langevin, BP 156, 38042 Grenoble Cedex 9, France*

[6]*Frontier Research Center, Tokyo Institute of Technology, 4259 Nagatsuta-cho, Midori-ku, Yokohama 226-8503, Japan*

[7]*Institut fuer Festkoerperforschung, Forschungszentrum Juelich, D-52425 Juelich, Germany*

We report the pressure and temperature dependence of the phonon density-of-states in superconducting Ca$_{0.6}$Na$_{0.4}$Fe$_2$As$_2$ (T$_c$=21 K) and the parent compound CaFe$_2$As$_2$, using inelastic neutron scattering. We observe no significant change in the phonon spectrum for Ca$_{0.6}$Na$_{0.4}$Fe$_2$As$_2$ at 295 K up to pressures of 5 kbar. The phonon spectrum for CaFe$_2$As$_2$ shows softening of the low-energy modes by about 1 meV when decreasing the temperature from 300 K to 180 K. There is no appreciable change in the phonon density of states across the structural and anti-ferromagnetic phase transition at 172 K. These results, combined with our earlier temperature dependent phonon density of states measurements for Ca$_{0.6}$Na$_{0.4}$Fe$_2$As$_2$, indicate that the softening of low-energy phonon modes in these compounds may be due to the interaction of phonons with electron or short-range spin fluctuations in the normal state of the superconducting compound as well as in the parent compound. The phonon spectra are analyzed with ab-initio and empirical model calculations giving partial densities of states and dispersion relations.

## I. Introduction

MFe$_2$As$_2$ compounds are known to be stable with divalent M = Ba, Ca, Sr, and Eu atoms. At room temperature these compounds crystallize in the ThCr$_2$Si$_2$-type tetragonal (I4/mmm) structure. They are, in many aspects, similar to the family of ROFeAs (R=Rare earth) compounds, which crystallize in the tetragonal ZrCuSiAs-type structure (P4/mmm). These compounds have recently attracted immense attention [1-19] in the scientific community. Both systems feature layers of FeAs. They undergo [9,10] a transition from tetragonal to orthorhombic symmetry below room temperature and order anti-ferromagnetically in the orthorhombic structure. Electron or hole doping into the parent compounds suppresses the structural and magnetic instabilities and induces superconductivity [2]. The magnetic order can be suppressed in the parent compound by the application of pressure [4-6]. Pressure can also induce superconductivity and CaFe$_2$As$_2$ is, in this context, the only compound that shows superconductivity [4] at a rather low pressure of 3.5 kbar. CaFe$_2$As$_2$ is thus an ideal candidate for the pressure dependent investigation of structural and magnetic properties by neutron scattering. High-pressure neutron diffraction measurements show [5,6] that CaFe$_2$As$_2$ undergoes a transition to a "collapsed tetragonal phase" under applied pressure below 50 K in which the c-parameter is reduced by 10 %. Band structure calculations show [5] that the collapse of the tetragonal phase is due to the loss of magnetism in the Fe system. This leads to further speculation that the loss of the iron magnetic moment may be important to stabilize the superconducting phase in iron-pnictide superconductors. Recent high-pressure inelastic neutron scattering measurements [11] of CaFe$_2$As$_2$ showed that the antiferromagnetic spin fluctuations observed in the ambient pressure paramagnetic, tetragonal phase are strongly suppressed in the high-pressure collapsed tetragonal phase.

Superconductivity in FeAs compounds is believed to be mediated by antiferromagnetic spin fluctuations. Keeping this in mind inelastic neutron scattering measurements have been carried out on polycrystalline Ba$_{0.6}$K$_{0.4}$Fe$_2$As$_2$ [13] as well as on single crystals of Ba(Fe$_{0.92}$Co$_{0.08}$)$_2$As$_2$ [14(a)] and BaFe$_{1.9}$Ni$_{0.10}$As$_2$ [14(b)], which indicate evidence of a resonant spin excitation. Phonon spectra have been shown to depend on the magnetic state [16] and a possible role in the pairing mechanism cannot be excluded. Earlier we have reported phonon dynamics for parent BaFe$_2$As$_2$ [17] as well as superconducting Sr$_{0.6}$K$_{0.4}$Fe$_2$As$_2$ and Ca$_{0.6}$Na$_{0.4}$Fe$_2$As$_2$ compounds [18] using the techniques of inelastic neutron scattering and lattice dynamics calculations. In this paper, we report the measurements of phonon spectra at high pressure for Ca$_{0.6}$Na$_{0.4}$Fe$_2$As$_2$ and the temperature dependence for the parent CaFe$_2$As$_2$. Lattice dynamical calculations are also carried out for CaFe$_2$As$_2$ using the shell model and ab-initio methods. The discussion given in this paper will be based on our earlier investigations of the BaFe$_2$As$_2$ and Ca$_{0.6}$Na$_{0.4}$Fe$_2$As$_2$ compounds. Section II gives an outline of the experimental technique, as adopted here. The details about the lattice dynamics calculations are given in Sec. III, followed by the results and discussion, and conclusions in Secs. IV and V, respectively.

**II. Experimental**

The polycrystalline samples of CaFe$_2$As$_2$ and Ca$_{0.6}$Na$_{0.4}$Fe$_2$As$_2$ (T$_c$ =21 K) were prepared by heating stoichiometric mixtures of the corresponding purified elements. Structural analysis from x-ray powder diffraction indicates that the CaFe$_2$As$_2$ sample contains about 2% of FeAs as an impurity phase. For the high-pressure, inelastic measurements on Ca$_{0.6}$Na$_{0.4}$Fe$_2$As$_2$ we have used the same sample as was used in our previous measurements [18] of the temperature dependence of the phonon density of states. The details about the characterization of superconducting Ca$_{0.6}$Na$_{0.4}$Fe$_2$As$_2$ are given in our previous publication [18]. The inelastic neutron scattering experiments were carried out using the IN4C and IN6 time-of-flight spectrometers at the Institut Laue Langevin (ILL), France. Both the spectrometers are based on the time-of-flight technique and are equipped with a large detector bank covering a wide range of about 10° to 115° of scattering angle. A polycrystalline sample of 10 gm of CaFe$_2$As$_2$ was placed inside a sealed aluminum container in the form of a thin slab, which was mounted in a cryostat, at 45° to the incident neutron beam, for temperature dependent measurements. For high pressure measurements we have used about 8 grams of Ca$_{0.6}$Na$_{0.4}$Fe$_2$As$_2$ sample.

An incident neutron wavelength of 1.2 Å (58.8 meV) was chosen for the IN4C measurements, which allowed the data collection in the neutron-energy loss mode. The measurements for CaFe$_2$As$_2$ were performed at 2 K, 140 K and 190 K. The high-resolution measurements for CaFe$_2$As$_2$ at 300 K and 180 K at ambient pressure and high pressure measurements at 295 K for superconducting Ca$_{0.6}$Na$_{0.4}$Fe$_2$As$_2$ were performed on IN6. For these measurements we have used an incident neutron wavelength of 5.1 Å (3.12 meV) in neutron energy gain mode. The elastic energy resolution was about 80 μeV. The high pressure measurements for Ca$_{0.6}$Na$_{0.4}$Fe$_2$As$_2$ at ambient pressure, 0.3 kbar, 2.8 kbar and 5 kbar, at 295 K were performed using a gas pressure cell with argon as the pressure transmitting medium. The incoherent approximation [20] has been used for extracting neutron weighted phonon density of states from the measured scattering function S(Q,E). The weighting factors $\frac{4\pi b_k^2}{m_k}$ for various atoms in the units of barns/amu are: Ca: 0.071; Fe: 0.208 and As: 0.073. The experimental one-phonon spectrum is obtained by subtracting the multi-phonon contribution from the experimental data. In the case of the IN6 data multi-phonon contributions were obtained via a self-consistent formalism, while Sjolander formalism [21] has been used for obtaining the multi-phonon contributions for the IN4C data.

## III. Lattice dynamical calculations

The phonon frequencies as a function of wave vectors in the entire Brillouin zone have been calculated for $CaFe_2As_2$ using quantum-mechanical ab-initio methods and semiempirical interatomic potentials as in [18]. The parameters of the interatomic potential satisfy the conditions of static and dynamic equilibrium [22,23]. The shell model calculations have been carried out using the current version of the code DISPR [24] developed at Trombay.

Ab-initio calculations were performed using the projector-augmented wave (PAW) formalism [25] of the Kohn-Sham DFT [26, 27] at the generalized gradient approximation level (GGA), implemented in the Vienna ab-initio simulation package (VASP) [28, 29]. The GGA was formulated by the Perdew-Burke-Ernzerhof (PBE) [30,31] density functional. The Gaussian broadening technique was adopted and all results are well converged with respect to k-mesh and energy cutoff for the plane wave expansion. Experimentally refined crystallographic data in the low- and high-temperature ranges corresponding to the orthorhombic and tetragonal phases, respectively, have been considered. These structures were used to calculate the GDOS and dispersion relations for the orthorhombic phase under the *Fmmm* space group (number 69) having the local point group symmetry $D_{23}^{2h}$, and for the tetragonal phase under the *I4/mmm* space group (number 139) having the local point group symmetry $D_{17}^{4h}$. In the ab-initio lattice dynamics calculations, in order to determine all inter-atomic force constants, the supercell approach has been adopted [32]. Therefore for both phases, the single cell was used to construct a (2*a, 2*b, c) supercell containing 16 formula-units (80 atoms), and (3*a, 3*b, c) supercell containing 18 formula-units (90 atoms) for the orthorhombic and tetragonal phases, respectively (*a* and *b* being the shorter cell axes). Total energies and inter-atomic forces were calculated for the 18 and 16 structures resulting from individual displacements of the three symmetry inequivalent atoms, along the three Cartesian directions (±x, ±y and ±z). The 15 phonon branches corresponding to the 5 atoms in the primitive cell were extracted in subsequent calculations using the Phonon software [33].

## IV. Results and discussion

**A. High-pressure inelastic neutron scattering measurements on $Ca_{0.6}Na_{0.4}Fe_2As_2$**

Recently we have reported an experimental phonon study [18] of the $Sr_{0.6}K_{0.4}Fe_2As_2$ and $Ca_{0.6}Na_{0.4}Fe_2As_2$ superconducting compounds using inelastic neutron scattering. In both compounds the low-energy phonon modes soften with temperature. In general phonon modes are expected to shift towards higher energies with a decrease of the unit cell volume induced by a decrease in temperature. We speculated that the softening of low-energy phonons might be due to electron-phonon coupling effects. To separate effects of temperature and cell volume, we have carried out high-pressure, inelastic neutron scattering experiments for superconducting $Ca_{0.6}Na_{0.4}Fe_2As_2$. The measured phonon spectra at ambient pressure, 0.3 kbar, 2.9 kbar and 5 kbar are shown in Fig. 1. The large contributions to the measured spectra from the pressure cell do not allow us to obtain the experimental phonon spectra beyond 14 meV. Our results indicate that the compression of the unit cell volume has no effect on the phonon spectrum within the explored range.

High-pressure studies of $Ca_{0.6}Na_{0.4}Fe_2As_2$ have not been reported. However, the calculated variation of volume with pressure for $CaFe_2As_2$ using ab-initio methods [12] gives bulk modulus values of 56.2 GPa and 81.6 GPa respectively for the tetragonal and collapsed tetragonal state of $CaFe_2As_2$. Using the above tetragonal phase value of the bulk modulus we estimate a reduction in unit cell volume of about 0.88 % on compression from ambient pressure to 5 kbar. The change in volume on compression is roughly the same as that found (~ 0.7 %) for $CaFe_2As_2$ [10(c)] on lowering the temperature from 300 K to 140 K. We assume that thermal expansion behavior is nearly the same in both parent and superconducting Ca compounds. The change in phonon energies with temperature is due to "implicit" as well as "explicit" anharmonicities. The implicit anharmonicity of phonons is due to the change of the unit cell volume. The explicit anharmonicity includes changes in phonon frequencies due to thermal effects, like electron-phonon interactions. Our measurements show that compression of the unit cell volume of $Ca_{0.6}Na_{0.4}Fe_2As_2$ to 5 kbar has no effect on the phonon spectra of $Ca_{0.6}Na_{0.4}Fe_2As_2$. The combination of our present high-pressure results with our earlier temperature dependent measurements [18] of the phonon spectra for $Ca_{0.6}Na_{0.4}Fe_2As_2$ allows us to state that the softening of the low-energy phonon modes on lowering of the temperature is due to explicit effects. These explicit effects may be due to interaction of phonons with electron or short-range spin fluctuations in the normal state of superconducting sample.

**B. Temperature dependence of phonon spectra for $CaFe_2As_2$**

We have also measured the temperature dependence of the phonon spectra (Fig. 2(a)) for the parent compound $CaFe_2As_2$. Previous measurements on $Ca_{0.6}Na_{0.4}Fe_2As_2$ revealed [18] the softening of the low-energy modes on cooling from 300 K to 140 K, while no effect was observed on heating from 2 K to 50 K across the superconducting transition temperature of 21 K. $CaFe_2As_2$ has an orthorhombic-to-tetragonal structural and antiferromagnetic phase transition at 172 K on heating. Now in order to investigate whether phonon softening is related to structural and magnetic phase transition or due to paramagnetic fluctuations above 172 K, we have measured high-resolution phonon spectra using IN6 spectra at 300 K and 180 K.

Measurements on IN6 are performed with small incident neutron energy of 3.12 meV in the neutron-energy gain mode, which does not give enough intensity at low temperatures. Therefore measurements using IN4C spectrometer are carried out at 2K, 140 K and 190 K. The high-resolution measurements using IN6 spectrometer clearly show (Fig. 2) that low-energy phonon modes soften by about 1 meV while cooling from 300 K to 180 K, while spectra obtained from IN4C show (Fig. 3) no temperature dependence when heating from 2 K to 190 K. It seems therefore that the orthorhombic to tetragonal phase transition [10(b)] at 172 K has no effect on the phonon spectra. As discussed above, the phonon softening observed in our temperature dependent DOS measurements may be due to short-range spin fluctuations in the paramagnetic state of $CaFe_2As_2$.

Recent measurements [19] of the temperature dependence of phonon dispersion relations in $CaFe_2As_2$ shows that the zone boundary, transverse acoustic mode of energy about 10.5 meV, with polarization (1-10) along (110), shows softening on approaching the phase transition temperature (172 K) from higher temperature. However, above the phase transition temperature of 172 K the energy of the TA mode again shifts backs to about 10.5 meV. In our density of states measurements we observe that modes around 10 meV are most affected by the temperature. It should be noted that the density of states measurements are mainly sensitive to zone boundary modes. Our measurements on the polycrystalline sample are thus fully consistent with the single crystal measurements.

For $CaFe_2As_2$, the high-frequency bands around 34 meV are found to be narrower at 180 K and shifted to higher energies as expected on compression of the unit cell volume by a decrease of the temperature from 300 K to 180 K. The temperature dependence of the phonon spectra for $Ca_{0.6}Na_{0.4}Fe_2As_2$ showed a similar behaviour.

$CaFe_2As_2$ orders antiferromagnetically [10(b)] in the orthorhombic structure below 172 K. The experimental Bose factor corrected S(Q,E) plots for $CaFe_2As_2$ at 2 K, 140 K and 190 K measured using

IN4C spectrometer are shown in Fig. 4. The experimental data show no signs of magnetic excitations in the attainable (Q, E) range of IN4C. However recent measurements carried out on powder samples of BaFe$_2$As$_2$ [15], using the MERLIN spectrometer at ISIS, show evidence of spin excitations in the antiferromagnetically ordered state of BaFe$_2$As$_2$. This may be due to the fact that the (Q, E) range attainable at IN4C is different from that of MERLIN at low Q values. Further investigations might be necessary before drawing any final conclusions concerning CaFe$_2$As$_2$.

**C. Comparison of phonon spectra for CaFe$_2$As$_2$, Ca$_{0.6}$Na$_{0.4}$Fe$_2$As$_2$ and BaFe$_2$As$_2$**

Now we compare the high-resolution phonon density of states measured for the Ca, CaNa and Ba compounds using IN6 at 300 K. On partial doping of Na at the Ca site, the modes up to 12 meV show no change, while those above 12 meV soften by about 1 meV. Both the Ca and CaNa compounds have nearly the same value of the lattice parameter $a$ (~3.89 Å), while the lattice parameter $c$ in CaFe$_2$As$_2$ and Ca$_{0.6}$Na$_{0.4}$Fe$_2$As$_2$ is 11.758 Å and 12.066 Å, respectively. The doped compound has slightly longer M-As and Fe-As bond lengths that would result in a softening of the phonon modes, as reported above.

The phonon spectra (Fig. 2(b)) of the parent BaFe$_2$As$_2$ and CaFe$_2$As$_2$ compounds below 25 meV differ significantly. Both compounds have nearly the same lattice parameter $a$, while the lattice parameter $c$ in the Ca compound (11.758 Å) is about 10 % smaller than that of the Ba compound (13.04 Å). Similarly the mass of Ba (m= 137.34 amu) is large in comparison to Ca (m=40.08 amu). We expect that both the mass effect and the contraction of unit cell should result in shifting the phonon modes to higher energies but we find that in the Ca compound the low-energy modes up to 22 meV are shifted towards lower energies.

In order to understand this difference, we have calculated the partial density of states (Fig. 5) of various atoms (Section IVD) in BaFe$_2$As$_2$ and CaFe$_2$As$_2$. We find that the vibrational modes due to Ca atoms in CaFe$_2$As$_2$ scale approximately with the Ba vibrations in BaFe$_2$As$_2$ in the ratio expected from the masses of the Ca and Ba atoms. These modes thus behave as expected. Further we notice that there is a substantial difference in the vibrations of Fe and As atoms in both these compounds. The calculated Fe and As vibrations are found to soften in the range below 22 meV by about 2 meV in the Ca compound in comparison to the Ba compound. This is contrary to our expectation. We would naively guess that the shorter Ca-As and Fe-As bond lengths result in a hardening of all the Fe and As vibrations. We find this trend only for the peak at 32 meV, which indeed moves to higher energies. The unexpected softening of the rest of the Fe and As vibrations are found to be responsible for the softening of the entire phonon spectra (Fig. 2(b)) below 22 meV as seen in $CaFe_2As_2$.

**D. Phonon calculations in $CaFe_2As_2$**

The calculated phonon spectra using the shell model is shown in Fig. 3. The calculated spectra compare reasonably well with the experimental data. However there are discrepancies between the calculated and experimental phonon spectra at energies around 20 meV. We have also calculated the phonon spectra for $CaFe_2As_2$ using ab-initio methods for both the orthorhombic (low-T) and tetragonal (high-T) phases. For the tetragonal phase the calculated spectral profile is good (Fig. 3). However the Fe-As stretching frequency is overestimated since the cell optimization prior to the force constant calculation results in a significant shortening (collapse) [5] of the c axis and, accordingly, the Fe-As bond. Imposing the experimental value of the c-axis gives better agreement for the highest frequency vibration.

The density of states in the orthorhombic phase was calculated with and without the observed magnetic ordering (Fig. 3). All the observed features are well reproduced computationally. Without magnetic ordering, as for the tetragonal phase, there is a significant structural perturbation in the c direction and the Fe-As stretching vibration frequency is overestimated. The calculated orthorhombic distortion is found to be strongest for the observed magnetic structure and matches the experimental value. In this case, the magnetic coupling between the Fe-As planes helps to reproduce the measured value of the c-axis and the As z-coordinate. Accordingly the spectral frequencies and intensities match better the experimental data at 2 K.

The ab-initio results for the orthorhombic structure with observed magnetic ordering are in better agreement (Fig. 3) with the inelastic neutron scattering data compared to the empirical shell model calculations and the published ab-initio calculations [16] for the Ba compound. In order to understand this improved result and in the context of the foregoing discussion of the effect of cation mass and cell parameters (Section IV C), it is essential to consider the partial density of states (pDOS) for each atomic species (Fig. 5). For Ca, which is the lightest atom in the $CaFe_2As_2$ compound, the pDOS is localized in the frequency range from 15 to 24 meV. This limited frequency range is indicative of the dominantly ionic nature of the Ca interaction with the Fe-As layer. Covalent bonding would result in the pods of the light Ca atom covering the same frequency range as the Fe and As atoms. The Ba pDOS is localized at lower frequency in the range from 8 to 12 meV since it is the heaviest atomic species in the BaFe$_2$As$_2$ compound. Ba has a mass 3.5 times greater than Ca which approximately accounts for the frequency shift, again suggesting ionic interactions between the cations and the Fe-As layers that are not significantly modified on changing the cation.

The shift of the cation pDOS to higher frequency results in changes of the Fe and As pDOS in precisely the frequency range of the Ca pDOS. In the Ba case there are prominent peaks at 14 and 18 meV, whereas in the Ca system there is a single, broad spectral peak centred at 16 meV. The high frequency tail of the Ca signal also corresponds to increased intensity in the Fe and As pDOS at 22 meV, the frequency at which a peak is measured [16] for BaFe$_2$As$_2$, but not calculated.

Above 24 meV, there are two notable differences between the Ca and Ba cases. Firstly, the Ba system is characterized by sharper features in the pDOS for Fe and As and, secondly, the high frequency cut-off for Ba is lower than that of the Ca system, as in the experimental data (Fig. 2(b)). The sharp features in the pDOS can be understood from the phonon dispersion relations (Fig. 6). Close inspection and comparison of the two sets of dispersion relations reveals broad similarities, but the Ba case is characterized by dispersion branches that are flatter over wide ranges of reciprocal space. The branches involving the Ca cation occur at higher frequency than those of Ba, which appears to perturb all higher frequency branches. Accordingly there is only one gap in the GDOS of CaFe$_2$As$_2$ at 30 meV, whereas the BaFe$_2$As$_2$ also has a calculated gap at 20 meV, which is not observed experimentally.

From a computational point of view, this analysis leads us to tentatively conclude that the Ca cation couples dynamically more strongly to the Fe-As layers, whereas the Ba cation is chemically harder (its iconicity is higher) and more decoupled, due in part to the significantly lower frequency of the Ba pDOS. The extent of this decoupling is overestimated in our calculations [16] as evidenced by the fact that we do not reproduce the measured peak at 21 meV in BaFe$_2$As$_2$.

V. Conclusions

We have reported measurements of the pressure as well as the temperature dependence of phonon spectra for Ca$_{0.6}$Na$_{0.4}$Fe$_2$As$_2$ superconducting and parent CaFe$_2$As$_2$ compounds, respectively. Our measurements show that the structural phase transition as a function of temperature appears to be irrelevant for the observed phonon softening in parent CaFe$_2$As$_2$. The tetragonal to orthorhombic phase transition is suppressed in the superconducting Ca$_{0.6}$Na$_{0.4}$Fe$_2$As$_2$ compound. Phonon softening in Ca$_{0.6}$Na$_{0.4}$Fe$_2$As$_2$ and CaFe$_2$As$_2$ is found only above 180 K, corresponding to the paramagnetic state of $CaFe_2As_2$. The combined study of pressure as well as temperature dependence of the phonon spectra thus indicates that the softening of low-energy phonon modes in these compounds may be due to the interaction of phonons with the short-range spin fluctuations in the paramagnetic state of $CaFe_2As_2$, or due to electron-phonon coupling in the superconducting $Ca_{0.6}Na_{0.4}Fe_2As_2$. The comparison of phonon spectra for Ba and Ca compounds show strong renormalization effects in the phonon spectra of these compounds, which cannot be simply explained by the lattice contraction and mass effect.

FIG. 1. (Color online) The experimental phonon spectra for $Ca_{0.6}Na_{0.4}Fe_2As_2$ as a function of pressure at a fixed temperature of 295 K.

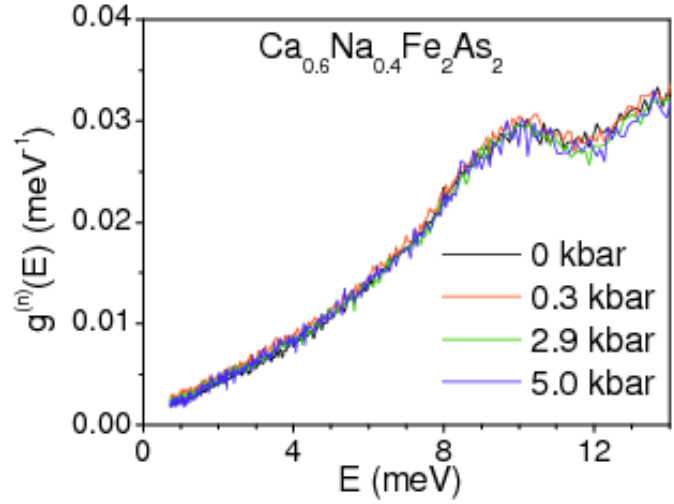

FIG. 2. (Color online) (a) The temperature dependence of the phonon spectra of $CaFe_2As_2$. (b) The comparison of the experimental phonon spectra for $CaFe_2As_2$, $Ca_{0.6}Na_{0.4}Fe_2As_2$ and $BaFe_2As_2$. The phonon spectra are measured with an incident neutron wavelength of 5.12 Å using the IN6 spectrometer at the ILL. The experimental phonon data for $BaFe_2As_2$ and $Ca_{0.6}Na_{0.4}Fe_2As_2$ are taken from Refs. [16] and [18], respectively. All the phonon spectra are normalized to unity.

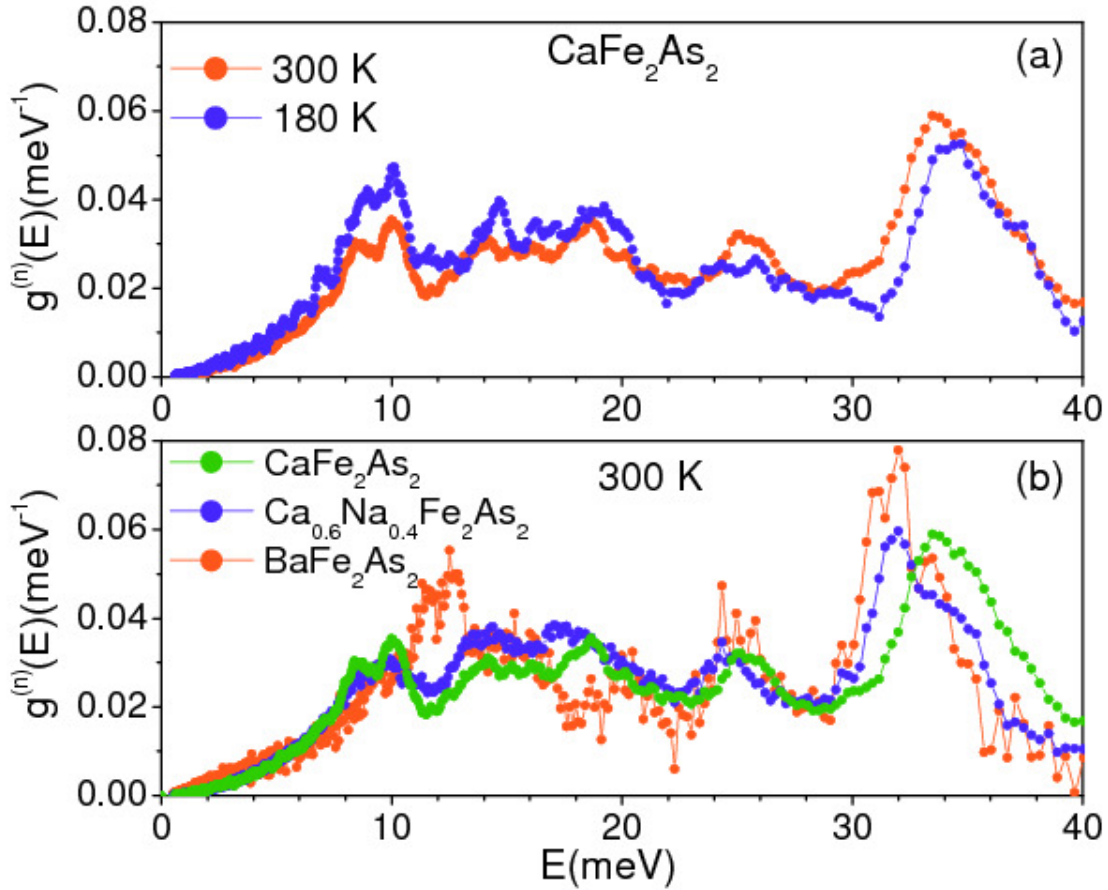

FIG. 3. (Color online) Comparison between the calculated and experimental phonon spectra of $CaFe_2As_2$. The measurements are carried out with incident neutron wavelength of 1.2 Å using the IN4C spectrometer at the ILL. For better visibility the phonon spectra are shifted along the y-axis by 0.02 $meV^{-1}$. The calculated spectra have been convoluted with a Gaussian of FWHM of 3 meV in order to describe the effect of energy resolution in the experiment.

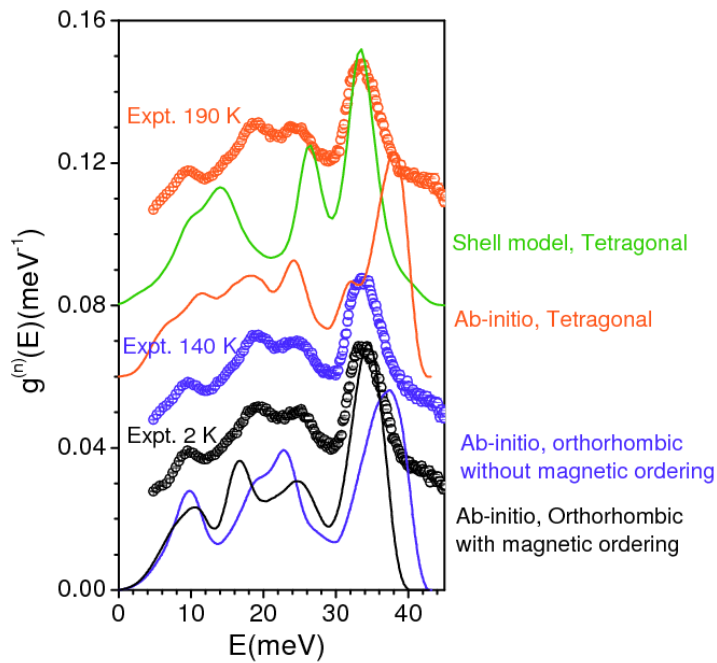

FIG. 4. (Color online) The experimental Bose factor corrected S(Q,E) plots for $CaFe_2As_2$ at 2 K, 140 K and 190 K measured using the IN4C spectrometer at the ILL with an incident neutron wavelength of 1.2 Å. The values of S(Q,E) are normalized to the mass of sample in the beam. For clarity, a logarithmic representation is used for the intensities.

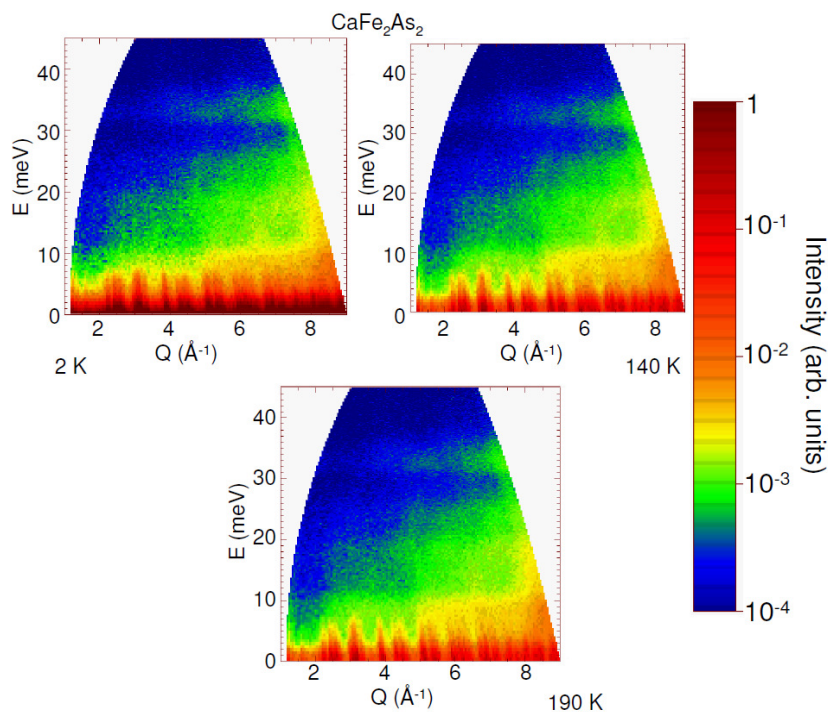

FIG. 5. (Color online) The calculated partial density of states for the various atoms in $CaFe_2As_2$ (red lines) and $BaFe_2As_2$ (black lines) from ab-initio calculations. The calculations are carried out for the orthorhombic phases with magnetic ordering. The partial density of states of various atoms and the total density of states are normalized to unity.

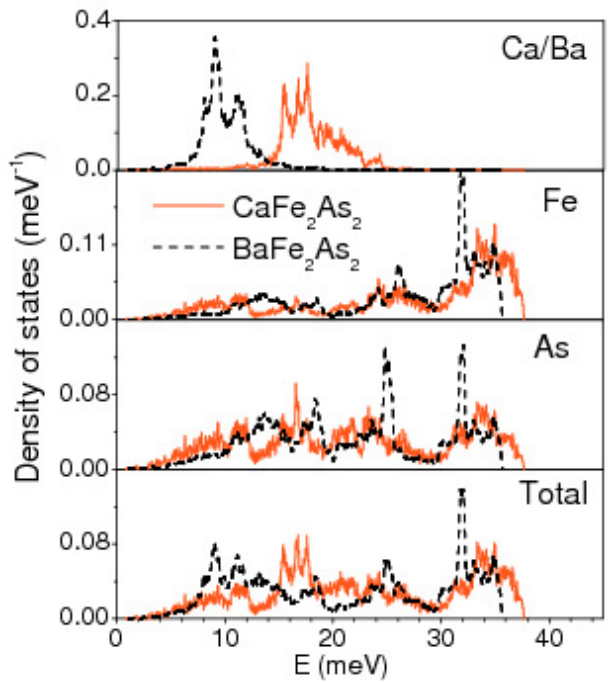

FIG. 6. The calculated phonon dispersion relations for $CaFe_2As_2$ (upper) and $BaFe_2As_2$ (lower) in the orthorhombic phases with magnetic ordering. The Bradley-Cracknell notation is used for the high symmetry points along which the dispersion relations are obtained: L = ( 1/2 , 0, 0), Y = ( 1/2 , 0, 1/2 ) and Z = ( 1/2 , 1/2 , 0).

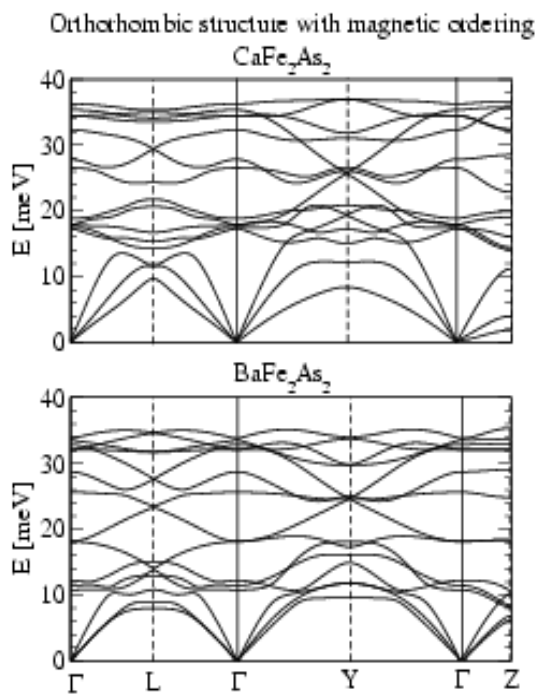